\documentclass[12pt,thmsa]{article}
\usepackage{amssymb}
\usepackage{amsfonts}
\usepackage{amsmath}
\usepackage{graphicx}
\usepackage{color}
\usepackage{hyperref}

\newtheorem{theorem}{Theorem}

\newtheorem{definition}{Definition}

\newtheorem{proposition}[theorem]{Proposition}
\newtheorem{lemma}[theorem]{Lemma}

\newenvironment{super-boxer}
{\begin{center}
\begin{tabular}{||p{0.95\columnwidth}||}
\hline \hline \\
}
{
\\ \\ \hline \hline
\end{tabular}
\end{center}
}
\definecolor{cadmiumred}{rgb}{0.00, 0.00, 0.00}
\input{tcilatex}
\begin{document}

\title{A Canon of Probabilistic Rationality\thanks{\color{cadmiumred}This
paper combines and supersedes two independent works: Paper 6 (Lindberg, 2012b) of the
thesis of Lindberg (2012a) and a working paper of Cerreia-Vioglio Maccheroni, Marinacci, and Rustichini
(2016). We thank Sean Horan, Marco Pavan (the
editor), an anonymous associate editor, and two anonymous referees for very
helpful comments, as well as the ERC (grants SDDM-TEA and INDIMACRO) and a
PRIN\ grant (2017CY2NCA) for financial support.}}
\author{Simone Cerreia-Vioglio$^{a}$, {Per Olov Lindberg}$^{b}$, Fabio
Maccheroni$^{a}$ \and Massimo Marinacci$^{a}$ and Aldo Rustichini$^{c}$ \\
$^{a}${\small Universit\`{a} Bocconi and Igier, }$^{b}${\small The Royal Institute of Technology}\\
$^{c}${\small University of Minnesota}}
\date{\today }
\maketitle

\begin{abstract}
We prove that a random choice rule satisfies Luce's Choice Axiom if and only
if its support is a choice correspondence that satisfies the Weak Axiom of
Revealed Preference, thus it consists of alternatives that are optimal
according to some preference, and random choice then occurs according to a
tie breaking among such alternatives that satisfies Renyi's Conditioning
Axiom.

Our result shows that the Choice Axiom is, in a precise formal sense, a
probabilistic version of the Weak Axiom. It thus supports Luce's view of his
own axiom as a \textquotedblleft canon of probabilistic
rationality.\textquotedblright
\end{abstract}

\newpage

\section{Introduction}

In 1977, twenty years after proposing it, Duncan Luce commented as follows
about his celebrated Choice Axiom:\footnote{%
Luce (1977, p. 229), emphasis added.}

\begin{quote}
\textquotedblleft Perhaps the greatest strength of the choice axiom, and one
reason it continues to be used, is as a \emph{canon of probabilistic
rationality}. It is a natural probabilistic formulation of K. J. Arrow's
famed principle of the \emph{independence of irrelevant alternatives}, and
as such it is a possible underpinning for rational, probabilistic theories
of social behavior.\textquotedblright
\end{quote}

This claim already appears in his 1957 and 1959 works that popularized the
axiom and the resulting stochastic choice model.\footnote{%
See Luce (1957, p. 6) and Luce (1959, p. 9).} The conceptual proximity of
Arrow's principle, typically identified with the set-theoretic version of
the Weak Axiom of Revealed Preference (WARP),\footnote{%
Arrow himself put forth this version of Samuelson's WARP in his 1948 and
1959 works.} and Luce's Choice Axiom is indeed often invoked. As well-known,
the former plays a key role in deterministic choice theory, the latter in
stochastic choice theory.

Yet, the formal relation between these two \emph{independence of irrelevant
alternatives (IIA)} assumptions has remained elusive so far.\footnote{%
See the discussion of Peters and Wakker (1991, p. 1789) and Wakker (2010, p.
373).} For instance, in analyzing several different IIA axioms Ray (1973)
writes:

\begin{quote}
\textquotedblleft Obviously IIA (Luce) falls in a different category
altogether [relative to IIA (Arrow)], being concerned with probabilistic
choices.\textquotedblright
\end{quote}

This note provides the missing link by showing that a random choice rule
satisfies Luce's Choice Axiom if and only if:

\begin{enumerate}
\item its support, the set of \emph{alternatives that can be chosen}, is a
rational choice correspondence \`a la Arrow (1948, 1959), so it consists of
alternatives that are optimal according to some preference;

\item tie-breaking among the optimal alternatives is consistent in the sense
of conditional probability \`a la Renyi (1955, 1956).
\end{enumerate}

In this way, our analysis formally supports Luce's \textquotedblleft
canonical rationality\textquotedblright \ claim for his Choice Axiom via a
lexicographic composition of deterministic rationality (WARP) and stochastic
consistency (Renyi's Conditioning Axiom).

\section{Preliminaries}

\subsection{Random choice rules\label{sect:rcr}}

Let $\mathcal{A}$ be the collection of all non-empty finite subsets of a
universal set $X$ of possible alternatives. The elements of $\mathcal{A}$
are called \emph{choice sets }and denoted by $A$, $B$ and $C$.

A map $\Gamma :\mathcal{A}\rightarrow \mathcal{A}$ such that $\Gamma \left(
A\right) \subseteq A$ for all choice sets $A$ is called \emph{choice
correspondence}. It is \emph{rational} when%
\begin{equation}
B\subseteq A\  \text{and }\Gamma \left( A\right) \cap B\neq \varnothing
\Longrightarrow \Gamma \left( B\right) =\Gamma \left( A\right) \cap B 
\tag{WARP}  \label{WARP}
\end{equation}%
This is the set-theoretic form of WARP considered by Arrow (1948, 1959). Its
IIA nature is best seen when $\Gamma $ is a function:%
\begin{equation*}
B\subseteq A\text{ and }\Gamma \left( A\right) \in B\Longrightarrow \Gamma
\left( B\right) =\Gamma \left( A\right)
\end{equation*}%
In words, adding suboptimal alternatives is irrelevant for choice behavior.

We denote by $\Delta \left( X\right) $ the set of all finitely supported
probability measures on $X$ and, for each $A\subseteq X$, by $\Delta \left(
A\right) $ the subset of $\Delta \left( X\right) $ consisting of the
measures assigning mass $1$ to $A$.

\begin{definition}
A \emph{random choice rule} is a function 
\begin{equation*}
\begin{array}{llll}
p: & \mathcal{A} & \rightarrow & \Delta \left( X\right) \\ 
& A & \mapsto & p_{A}%
\end{array}%
\end{equation*}%
such that $p_{A}\in \Delta \left( A\right) $ for all $A\in \mathcal{A}$.
\end{definition}

Given any alternative $a\in A$, we interpret $p_{A}\left( \left \{
a\right
\} \right) $, also denoted by $p\left( a,A\right) $, as the
probability that an agent chooses $a$ when the set of available alternatives
is $A$. More generally, if $B$ is a subset of $A$, we denote by $p_{A}\left(
B\right) $ or $p\left( B,A\right) $ the probability that the selected
element lies in $B $.\footnote{%
Formally, $x\mapsto p\left( x,A\right) $ for all $x\in X$ is the discrete
density of $p_{A}$, but with an abuse of notation $p_{A}\left( \cdot \right)
\ $is identified$\ $with $p\left( \cdot ,A\right) $; we also write $%
p_{A}\left( a\right) $ instead of $p_{A}\left( \left \{ a\right \} \right) $.%
} This probability can be viewed as the frequency with which an element in $%
B $ is chosen. In particular, the set of \emph{alternatives that can be
chosen }from $A$ is the \emph{support} of $p_{A}$, given by%
\begin{equation*}
\limfunc{supp}p_{A}=\left \{ a\in X:p\left( a,A\right) >0\right \}
\end{equation*}%
The condition $p_{A}\left( A\right) =1$ guarantees that it is a non-empty
subset of $A$, so that the \emph{support correspondence}%
\begin{equation*}
\begin{array}{cccc}
\limfunc{supp}p: & \mathcal{A} & \rightarrow & \mathcal{A} \\ 
& A & \mapsto & \limfunc{supp}p_{A}%
\end{array}%
\end{equation*}%
is a choice correspondence.

Finally, the standard way of comparing the probabilities of choices in two
different sets $B$ and $C$ are the \emph{odds} in favor of $B$ over $C$,
that is,%
\begin{equation*}
r_{A}\left( B,C\right) =\frac{p_{A}\left( B\right) }{p_{A}\left( C\right) }=%
\frac{\# \text{ of times an element in }B\text{ is chosen}}{\# \text{ of
times an element in }C\text{ is chosen}}
\end{equation*}%
for all $B,C\subseteq A$. As usual, given any $b$ and $c$ in $X$, we set $%
p\left( b,c\right) =p\left( b,\left \{ b,c\right \} \right) $ and%
\begin{equation*}
r\left( b,c\right) =\frac{p\left( b,c\right) }{p\left( c,b\right) }
\end{equation*}

\subsection{Luce's model}

The classical assumptions of Luce (1959) on $p$ are:\bigskip

\noindent \textbf{Positivity}\emph{\ }$p\left( a,b\right) >0$\emph{\ for all 
}$a,b\in X$\emph{.}\bigskip

\noindent \textbf{Choice Axiom }$p\left( a,A\right) =p\left( a,B\right)
p\left( B,A\right) $\emph{\ for all }$B\subseteq A$\emph{\ in }$\mathcal{A}$%
\emph{\ and all }$a\in B$\emph{.}\bigskip

The latter axiom says that the probability of choosing an alternative $a$
from the choice set $A$ is the probability of first selecting $B$ from $A$,
then choosing $a$ from $B$ (provided $a$ belongs to $B$). As observed by
Luce, formally this assumption corresponds to the fact that $\left \{
p_{A}:A\in \mathcal{A}\right \} $ is a conditional probability system in the
sense of Renyi (1955, 1956).\footnote{%
See Lemma 2 of Luce (1959) and Lemma \ref{lem:Luce_4} in the appendix. For
bibliographic accuracy, we remark that here we consider the Choice Axiom in
the form stated by Luce (1957) as Axiom 1. Under Positivity, this version
coincides with Axiom 1 of the 1959 book, and is the version later analyzed
by Luce himself in the retrospective of 1977.} Remarkably, Luce's Choice
Axiom is also equivalent to:\bigskip

\noindent \textbf{Odds Independence}%
\begin{equation}
\frac{p\left( a,b\right) }{p\left( b,a\right) }=\frac{p\left( a,A\right) }{%
p\left( b,A\right) }  \tag{OI}  \label{IIA}
\end{equation}%
\emph{for all }$A\in \mathcal{A}$\emph{\ and all }$a,b\in A$\emph{\ such
that }$p\left( a,A\right) /p\left( b,A\right) $\emph{\ is well defined.}%
\footnote{%
That is, different from $0/0$. See Lemma 3 of Luce (1959) when Positivity
holds and Lemma \ref{lem:Luce_4} in the appendix for the general case.}%
\bigskip

This axiom says that the odds for $a$ against $b$ are independent of the
other available alternatives.\footnote{%
For this reason, also this axiom often goes under the IIA name. To avoid
confusion, we use a less popular label.}

\begin{theorem}[Luce]
\label{thm:luce}A random choice rule $p:\mathcal{A}\rightarrow \Delta \left(
X\right) $ satisfies Positivity and the Choice Axiom if and only if there
exists $\alpha :X\rightarrow \mathbb{R}$ such that 
\begin{equation}
p\left( a,A\right) =\dfrac{e^{\alpha \left( a\right) }}{\sum_{b\in
A}e^{\alpha \left( b\right) }}  \tag{LM}  \label{eq:LM}
\end{equation}%
for all $A\in \mathcal{A}$ and all $a\in A$.
\end{theorem}

This fundamental result in random choice theory also shows that, under the
Choice Axiom, Positivity is equivalent to the stronger assumption that $%
p_{A} $ has full support for all choice sets $A$.\bigskip

\noindent \textbf{Full Support}\emph{\ }$\limfunc{supp}p_{A}=A$\emph{\ for
all }$A\in \mathcal{A}$\emph{.}\bigskip

From a choice-theoretic perspective, this axiom is unduly restrictive and
may permit the choice of \textquotedblleft dominated\textquotedblright \
actions. This note shows what happens when removing from the Luce analysis
this extra baggage.

Finally, when $X$ is a separable metric space we may introduce a continuity
axiom.\bigskip

\noindent \textbf{Continuity}\emph{\ Given any }$x,y\in X$,\emph{\ if }$%
\left \{ x_{n}\right \} _{n\in \mathbb{N}}$\emph{\ converges to }$x$\emph{,
then}%
\begin{align*}
p\left( x_{n},y\right) & >0\text{\emph{\ for all }}n\in \mathbb{N}\text{ }%
\implies p\left( x,y\right) >0 \\
p\left( y,x_{n}\right) & >0\text{\emph{\ for all }}n\in \mathbb{N}\text{ }%
\implies p\left( y,x\right) >0
\end{align*}

This axiom has a natural interpretation: if, eventually, $x_{n}$ may be
always chosen (rejected) over $y$, and $x_{n}$ converges to $x$, then $x$
can be chosen (rejected) over $y$. Continuity is automatically satisfied
under Full Support as well as when $X$ is countable and endowed with the
discrete metric.

\section{Main result\label{sect:main}}

The next result generalizes Luce's Theorem \ref{thm:luce} by getting rid of
the Full Support assumption.

\begin{theorem}
\label{thm:main}The following conditions are equivalent for a random choice
rule $p:\mathcal{A}\rightarrow \Delta \left( X\right) $:

\begin{enumerate}
\item[(i)] $p$ satisfies the Choice Axiom;

\item[(ii)] there exist a function $\alpha :X\rightarrow \mathbb{R}$ and a
rational choice correspondence $\Gamma :\mathcal{A}\rightarrow \mathcal{A}$
such that 
\begin{equation}
p\left( a,A\right) =\left \{ 
\begin{array}{ll}
\dfrac{e^{\alpha \left( a\right) }}{\sum_{b\in \Gamma \left( A\right)
}e^{\alpha \left( b\right) }}\medskip & \qquad \text{if }a\in \Gamma \left(
A\right) \\ 
0 & \qquad \text{else}%
\end{array}%
\right.  \tag{CA}  \label{eq:pomp}
\end{equation}%
for all $A\in \mathcal{A}$ and all $a\in A$.
\end{enumerate}

In this case, $\Gamma $ is unique and given by $\Gamma \left( A\right) =%
\limfunc{supp}p_{A}$ for all $A\in \mathcal{A}$.
\end{theorem}

Since $\Gamma $ is a rational choice correspondence, the relation $\succ $
defined by%
\begin{equation*}
a\succ b\iff a\neq b\text{ and }\Gamma \left( \left \{ a,b\right \} \right)
=\left \{ a\right \} \iff b\notin \Gamma \left( \left \{ a,b\right \} \right)
\end{equation*}%
is a strict preference (see Kreps, 1988) and the corresponding weak
preference 
\begin{equation*}
b\succsim a\iff a\nsucc b\iff b\in \Gamma \left( \left \{ a,b\right \}
\right) \iff p\left( b,a\right) >0
\end{equation*}%
is such that $\Gamma \left( A\right) =\left \{ a\in A:a\succsim b\text{ for
all }b\in A\right \} $.

When $X$ is countable, $\succsim $ is automatically represented by a utility
function $u$ and so we have%
\begin{equation*}
\Gamma \left( A\right) =\arg \max_{a\in A}u\left( a\right)
\end{equation*}%
In general, some additional conditions are needed, as next we show.

\begin{proposition}
\label{prop:u}If $X$ is a separable metric space, then the random choice
rule $p$ in Theorem \ref{thm:main} satisfies Continuity if and only if there
exists a continuous $u:X\rightarrow \mathbb{R}$ such that $\Gamma \left(
A\right) =\arg \max_{a\in A}u\left( a\right) $ for all $A\in \mathcal{A}$.
\end{proposition}

A two-stage decision process appears in formula (\ref{eq:pomp}): first
rational selection from the choice set $A$ via maximization of preference $%
\succsim $ (or utility $u$), then Lucean tie-breaking to choose among the
optimal alternatives.

While the optimization structure of the first stage is clear, more can be
said about the tie-breaking structure of the second stage in that Theorem %
\ref{thm:main} describes only its functional form. To this end, recall that
a random choice rule $p$ is based on a \emph{Random Preference Model} if
there is a (measurable) collection 
\begin{equation*}
\left \{ \succ _{\omega }\right \} _{\omega \in \left( \Omega ,\mathcal{F}%
,\Pr \right) }
\end{equation*}%
of strict preferences such that, for all $a\in A\in \mathcal{A}$, 
\begin{equation*}
p\left( a,A\right) =\Pr \left( \omega \in \Omega :a\succ _{\omega }b\text{%
\quad }\forall b\in A\setminus \left \{ a\right \} \right)
\end{equation*}%
In particular, a Random Preference Model is \emph{Lucean} if $p\left( \cdot
,A\right) $ has the Luce form (\ref{eq:LM}).

A piece of terminology: the \emph{lexicographic composition} of two binary
relations $\succ $ and $\succ ^{\prime }$ is the binary relation $\succ
\circ \succ ^{\prime }$ defined by%
\begin{equation*}
a\succ \circ \succ ^{\prime }b\iff a\succ b\text{ or }a\sim b\text{ and }%
a\succ ^{\prime }b
\end{equation*}%
For instance, $>_{1}\circ >_{2}$ is the usual lexicographic preference on
the Cartesian plane.\footnote{%
Here $>_{i}$ is defined by $\left. \left( a_{1},a_{2}\right) >_{i}\left(
b_{1},b_{2}\right) \Longleftrightarrow a_{i}>b_{i}.\right. $}

We can now state the announced characterization.

\begin{proposition}
\label{prop:lexi}The following conditions are equivalent for a random choice
rule $p:\mathcal{A}\rightarrow \Delta \left( X\right) $:

\begin{enumerate}
\item[(i)] $p$ satisfies the Choice Axiom;

\item[(ii)] $\limfunc{supp}p:\mathcal{A}\rightarrow \mathcal{A}$ is a
rational choice correspondence and%
\begin{equation}
p_{B}\left( a\right) =\frac{p_{A}\left( a\right) }{p_{A}\left( B\right) } 
\tag{COND}  \label{eq:cond}
\end{equation}%
for all $B\subseteq A\in \mathcal{A}$ and all $a\in B\cap \limfunc{supp}%
p_{A} $.

\item[(iii)] there exist a strict preference $\succ $ on $X$ and a Lucean
Random Preference Model $\left \{ \succ _{\omega }\right \} _{\omega \in
\left( \Omega ,\mathcal{F},\Pr \right) }$ such that $p$ is based on the
lexicographic Random Preference Model 
\begin{equation*}
\left \{ \succ \circ \succ _{\omega }\right \} _{\omega \in \left( \Omega ,%
\mathcal{F},\Pr \right) }
\end{equation*}
\end{enumerate}
\end{proposition}

This result presents two \textquotedblleft deconstructions\textquotedblright
\ of the Choice Axiom that both shed light on the second tie-breaking stage
in (\ref{eq:pomp}).

Specifically, to interpret (ii) observe that WARP\ says that, if $a$ can be
chosen from $A$ (i.e., $p_{A}\left( a\right) >0$) and belongs to $B\subseteq
A$, then it can be chosen also from $B$. But, this axiom is silent about the
relation between the frequencies of choice in the two sets $A$ and $B$.
Formula (\ref{eq:cond}) requires them to be related by the Conditioning
Axiom of Renyi (1955, 1956), a classical probabilistic consistency
condition. In particular, (\ref{eq:cond}) per se is weaker than Luce's
Choice Axiom, which imposes $p_{A}\left( a\right) =p_{B}\left( a\right)
p_{A}\left( B\right) $ for all $a\in B\subseteq A$, not just for the
elements $a$ in $B$ that can be chosen from $A$.

To interpret (iii), note that the first-stage preference $\succ $ determines
the support of $p$, while the second stage Random Preference Model $\left \{
\succ _{\omega }\right \} _{\omega \in \left( \Omega ,\mathcal{F},\Pr
\right) }$ is the formal description of the Lucean tie-breaking among
optimizers that we previously discussed.

Finally, (iii) also says that, when $X$ is countable, random choice rules
that satisfy the Choice Axiom are random utility models {\color{cadmiumred}%
(RUM)}, something not obvious from the definition.\footnote{%
\color{cadmiumred}See Section \ref{sect:irum} below for an independent RUM
representation.} This opens the way to the study of general compositions of
strict preferences and random utility models. The object of current
research, a such study goes beyond the scope of this note.

\section{Remarks}

\begin{enumerate}
\item By considering a random choice rule $p_{A}$ to describe the frequency
with which elements are chosen from $A$, we make the standard interpretation
of the choice correspondence $\Gamma \left( A\right) =\limfunc{supp}p_{A}$
as the the set of\emph{\ alternatives that can be chosen from }$A$ (cf. Sen,
1993) operational and formally meaningful. Here, \textquotedblleft can be
chosen\textquotedblright \ means chosen with positive frequency.

\item The second stage of randomization, disciplined by $\alpha $, can be
interpreted in the spirit of Salant and Rubinstein (2008) as capturing
observable information which is irrelevant in the rational assessment of the
alternatives, but nonetheless affects choice and may reveal how previous
experiences and mental associations affect the selection from the optimal $%
\Gamma \left( A\right) $.

\item The distinct roles of $u$ and $\alpha $ become clear once our result
is related to the random utility representation of the Luce model. In fact, $%
u$ corresponds to the \emph{systematic} component of the agent \emph{utility}%
, and $\alpha $ to the \emph{alternative-specific bias} in the Multinomial
Logit Model.\footnote{%
See the seminal McFadden (1973) as well as Ben Akiva and Lerman (1985) and
Train (2009) for textbook treatments.} Specifically, Theorem \ref{thm:main}
shows that a random choice rule $p$ has the form (\ref{eq:pomp}) if and only
if, given any $A\in \mathcal{A}$ and any $a\in A$,%
\begin{equation*}
p_{A}\left( a\right) =\lim_{\lambda \rightarrow 0}\Pr \left( \omega \in
\Omega :u\left( a\right) +\lambda \epsilon _{a}\left( \omega \right)
>u\left( b\right) +\lambda \epsilon _{b}\left( \omega \right) \text{\quad }%
\forall b\in A\setminus \left \{ a\right \} \right)
\end{equation*}%
where $u$ is a utility function that rationalizes $\Gamma $, $\left \{
\epsilon _{x}\right \} _{x\in X}$ is a collection of independent errors with
type I extreme value distribution, specific\emph{\ }mean $\alpha \left(
a\right) $, common variance $\pi ^{2}/6$, and $\lambda $ is the noise level.
In fact, {\footnotesize 
\begin{eqnarray*}
&&\lim_{\lambda \rightarrow 0}\Pr \left( \omega \in \Omega :u\left( a\right)
+\lambda \epsilon _{a}\left( \omega \right) >u\left( b\right) +\lambda
\epsilon _{b}\left( \omega \right) \text{\quad }\forall b\in A\setminus
\left \{ a\right \} \right) \\
&=&\lim_{\lambda \rightarrow 0}\Pr \left( \omega :\frac{u\left( a\right) }{%
\lambda }+\alpha \left( a\right) +\left[ \epsilon _{a}\left( \omega \right)
-\alpha \left( a\right) \right] >\frac{u\left( b\right) }{\lambda }+\alpha
\left( b\right) +\left[ \epsilon _{b}\left( \omega \right) -\alpha \left(
b\right) \right] \  \forall b\neq a \right) \\
&=&\lim_{\lambda \rightarrow 0}\dfrac{e^{\frac{u\left( a\right) }{\lambda }%
+\alpha \left( a\right) }}{\sum_{b\in A}e^{\frac{u\left( b\right) }{\lambda }%
+\alpha \left( b\right) }}=\dfrac{e^{\alpha \left( a\right) }}{\sum_{b\in
\arg \max_{A}u}e^{\alpha \left( b\right) }}\delta _{a}\left( \arg \max
\nolimits_{A}u\right) =p_{A}\left( a\right)
\end{eqnarray*}
} \vspace*{-20pt}

Our analysis thus shows that, when noise vanishes, optimal choice is
governed by $u$ and tie-breaking among optimal alternatives is
stochastically driven by alternative-specific biases captured by $\alpha $.

\item A similar interpretation arises when adopting the perspective of
Matejka and McKay (2015) on the Multinomial Logit Model as the outcome of an
optimal information acquisition problem. In this case, $u$ is the true
(initially unknown) payoff of alternatives, $\alpha $ captures a prior
belief on payoffs held before engaging in experimentation, and $\lambda $ is
the cost of one unit of information.

Here our analysis shows that, when the cost of information vanishes, optimal
alternatives are selected without error, and prior beliefs only govern the
tie-breaking among such alternatives.
\end{enumerate}

\section{Related literature}

The study of the relations between axiomatic decision theory and stochastic
choice has been recently an active field of research. Horan (2020) and Ok
and Tserenjigmid (2020) are the most recent works that we are aware of. The
former also provides an insightful review of the state of the art. The
latter expands on the main conceptual topic of this note: the relation
between deterministic and probabilistic \textquotedblleft
rationality.\textquotedblright

Horan (2020) axiomatically unifies Luce (1956, 1959) in a random choice
model of imperfect discrimination of the form 
\begin{equation}
p\left( a,A\right) =\left \{ 
\begin{array}{ll}
\dfrac{e^{\alpha \left( a\right) }}{\sum_{b\in \Gamma \left( A\right)
}e^{\alpha \left( b\right) }}\medskip & \qquad \text{if }a\in \Gamma \left(
A\right) \\ 
0 & \qquad \text{else}%
\end{array}%
\right.  \tag{GLM}  \label{eq:GLM}
\end{equation}%
where $\Gamma $ is a utility correspondence based on $\alpha $.
Specifically, in Horan, $\Gamma $ describes the degree of imperfection in
the discrimination of the $\alpha $-values of alternatives; on the contrary,
in this note $\alpha $ and $\Gamma $ are independent, with the former
tie-breaking the optimizers identified by the latter.

Horan also compares and provides alternative axiomatizations of several
\textquotedblleft General Luce Models\textquotedblright \ (the name is of
Echenique and Saito, 2019) of the form (\ref{eq:GLM}), which correspond to
different specifications of the properties of $\Gamma $: Ahumada and Ulku
(2019), Dogan and Yildiz (2019), Echenique and Saito (2019), and McCausland
(2009).

In particular, Dogan and Yildiz (2019) and Horan (2020) provide alternative
characterizations of (\ref{eq:pomp}): the former based on supermodularity of
odds, the latter on the product rule and a transitivity condition of
Fishburn (1978). These results ---together with our characterizations of (%
\ref{eq:pomp}) through the Choice Axiom alone, or WARP and conditioning---
provide a full perspective on \textquotedblleft rational
choice\textquotedblright \ followed by \textquotedblleft rational
tie-breaking.\textquotedblright

Like us, Ok and Tserenjigmid (2020) regard the support of a random choice
rule as a deterministic choice correspondence, and they analyze its
rationality properties for several different random choice rules. Following
Fishburn (1978), they also consider the entire family of deterministic
choice correspondences that lie between the support of $p$ and its subset
consisting of the alternatives that are chosen with highest frequency
(rather than with positive frequency).

\appendix

\section{Proofs and related analysis}

\subsection{Independent RUM\ representations\label{sect:irum}}

At the end of Section \ref{sect:main}, we observed how Proposition \ref%
{prop:lexi}.(iii) shows that, when $X$ is countable, random choice rules
that satisfy the Choice Axiom are random utility models. Here we expand on
this topic by providing an explicit independent random utility
representation for the random choice rule (\ref{eq:pomp}) of Theorem \ref%
{thm:main}, which holds whenever $\Gamma $ is the \textquotedblleft $\arg
\max $\textquotedblright \ of a utility function $u:X\rightarrow \mathbb{R}$
with discrete range. Note that, while this requires $u\left( X\right) $ to
be countable, no assumption is made on the cardinality of $X$.

\begin{proposition}
Let $u,\alpha :X\rightarrow \mathbb{R}$ and consider the random choice rule $%
p:\mathcal{A}\rightarrow \Delta \left( X\right) $ defined by 
\begin{equation*}
p\left( a,A\right) =\left \{ 
\begin{array}{ll}
\dfrac{e^{\alpha \left( a\right) }}{\sum_{b\in \arg \max_{z\in A}u\left(
z\right) }e^{\alpha \left( b\right) }}\medskip & \qquad \text{if }a\in \arg
\max_{z\in A}u\left( z\right) \\ 
0 & \qquad \text{else}%
\end{array}
\right.
\end{equation*}
for all $A\in \mathcal{A}$ and all $a\in A$. If $u\left( X\right) $ is a
discrete subset of $\mathbb{R}$, then there exists a collection $\left \{
U_{x}\right \} _{x\in X}$ of independent random variables such that 
\begin{equation*}
p\left( a,A\right) =\Pr \left( \omega \in \Omega :U_{a}\left( \omega \right)
>U_{b}\left( \omega \right) \text{\quad }\forall b\in A\setminus \left \{
a\right \} \right)
\end{equation*}
for all $A\in \mathcal{A}$ and all $a\in A$.
\end{proposition}

\noindent \textbf{Proof} Let $\left \{ V_{x}\right \} _{x\in X}$ be a
collection of independent random variables such that%
\begin{equation*}
\dfrac{e^{\alpha \left( a\right) }}{\sum_{b\in A}e^{\alpha \left( b\right) }}
=\Pr \left( \omega \in \Omega :V_{a}\left( \omega \right) >V_{b}\left(
\omega \right) \text{\quad }\forall b\in A\setminus \left \{ a\right \}
\right)
\end{equation*}%
for all $A\in \mathcal{A}$ and all $a\in A$, and assume that $-1<V_{x}\left(
\omega \right) <1$ for all $x\in X$ and all $\omega \in \Omega $.\footnote{%
This is without loss of generality because one can always take the
representation of McFadden (1973) and apply an arctangent transformation.}
Since $u\left( X\right) $ is discrete, for each $x\in X$ there exists a
constant $r_{x}>0$ which only depends on $u\left( x\right) $ such that 
\begin{equation*}
u\left( x\right) >u\left( y\right) \implies u\left( x\right) -r_{x}>u\left(
y\right) +r_{y}
\end{equation*}%
Define $U_{x}=u\left( x\right) +r_{x}V_{x}$ and note that $\left \{
U_{x}\right \} _{x\in X}$ is a collection of independent random variables
too.

Now arbitrarily choose $A\in \mathcal{A}$ and set $B=\arg
\max_{z\in A}u\left( z\right) $ and $C=A\setminus B$. Two cases have to be
considered.

If $a\in B$, then 
\begin{equation*}
p\left( a,A\right) =\dfrac{e^{\alpha \left( a\right) }}{\sum_{b\in
B}e^{\alpha \left( b\right) }}=\Pr \left( \omega \in \Omega :V_{a}\left(
\omega \right) >V_{b}\left( \omega \right) \text{\quad }\forall b\in
B\setminus \left \{ a\right \} \right)
\end{equation*}%
since $u\left( b\right) =u\left( a\right) $ for all $b\in B$, then $%
r_{b}=r_{a}$ for all $b\in B$, thus 
\begin{eqnarray*}
p\left( a,A\right) &=&\Pr \left( \omega \in \Omega :u\left( a\right)
+r_{a}V_{a}\left( \omega \right) >u\left( b\right) +r_{b}V_{b}\left( \omega
\right) \text{\quad }\forall b\in B\setminus \left \{ a\right \} \right) \\
&=&\Pr \left( \omega \in \Omega :U_{a}\left( \omega \right) >U_{b}\left(
\omega \right) \text{\quad }\forall b\in B\setminus \left \{ a\right \}
\right)
\end{eqnarray*}%
But, for all $c\in C=A\setminus B$ and all $\omega \in \Omega $,%
\begin{equation*}
U_{c}\left( \omega \right) =u\left( c\right) +r_{c}V_{c}\left( \omega
\right) <u\left( c\right) +r_{c}
\end{equation*}%
and $u\left( a\right) >u\left( c\right) $ implies $u\left( a\right)
-r_{a}>u\left( c\right) +r_{c}$, hence%
\begin{equation*}
U_{a}\left( \omega \right) =u\left( a\right) +r_{a}V_{a}\left( \omega
\right) >u\left( a\right) -r_{a}>u\left( c\right) +r_{c}>U_{c}\left( \omega
\right)
\end{equation*}%
Thus, $U_{a}\left( \omega \right) >U_{c}\left( \omega \right) $ for all $%
c\in C$ and all $\omega \in \Omega $, so 
\begin{eqnarray*}
p\left( a,A\right) &=&\Pr \left( \omega \in \Omega :U_{a}\left( \omega
\right) >U_{b}\left( \omega \right) \text{\quad }\forall b\in B\setminus
\left \{ a\right \} \text{ and }\forall b\in C\right) \\
&=&\Pr \left( \omega \in \Omega :U_{a}\left( \omega \right) >U_{b}\left(
\omega \right) \text{\quad }\forall b\in A\setminus \left \{ a\right \}
\right)
\end{eqnarray*}

If instead $c\in C$, then taking $a\in B$ as above, $U_{a}\left( \omega
\right) >U_{c}\left( \omega \right) $ for all $\omega \in \Omega $, then 
\begin{eqnarray*}
0 &=&\Pr \left( \omega \in \Omega :U_{c}\left( \omega \right) >U_{a}\left(
\omega \right) \right) \\
&\geq &\Pr \left( \omega \in \Omega :U_{c}\left( \omega \right) >U_{b}\left(
\omega \right) \text{\quad }\forall b\in A\setminus \left \{ c\right \}
\right)
\end{eqnarray*}%
whence%
\begin{equation*}
p\left( c,A\right) =0=\Pr \left( \omega \in \Omega :U_{c}\left( \omega
\right) >U_{b}\left( \omega \right) \text{\quad }\forall b\in A\setminus
\left \{ c\right \} \right)
\end{equation*}%
as wanted.\hfill $\blacksquare $

\subsection{Proofs}

A preference on $X$ can be given in either strict form, $\succ $, or weak
form, $\succsim $.

\begin{itemize}
\item In the first case, $\succ $ is required to be asymmetric and
negatively transitive, and $\succsim $ is defined by%
\begin{equation}
a\succsim b\text{ if and only if }\lnot \left( b\succ a\right)
\label{eq:weak}
\end{equation}

\item In the second case, $\succsim $ is required to be complete and
transitive, and $\succ $ is defined by%
\begin{equation}
b\succ a\text{ if and only if }\lnot \left( a\succsim b\right)
\label{eq:strict}
\end{equation}
\end{itemize}

These approaches are well known to be interchangeable,\footnote{%
See Kreps (1988, p. 11).} and for this reason we call \emph{weak order} both 
$\succ $ and $\succsim $ with the understanding that they are related by the
equivalent (\ref{eq:weak}) or (\ref{eq:strict}).

\begin{lemma}
\label{lem:Luce_4}Let $p:\mathcal{A}\rightarrow \Delta \left( X\right) $ be
a random choice rule. The following conditions are equivalent:

\begin{enumerate}
\item[(i)] $p$ is such that, $p_{A}\left( C\right) =p_{B}\left( C\right)
p_{A}\left( B\right) $\ for all $C\subseteq B\subseteq A$ in $\mathcal{A}$;

\item[(ii)] $p$ satisfies the Choice Axiom;

\item[(iii)] $p$ is such that $p\left( b,B\right) p\left( a,A\right)
=p\left( a,B\right) p\left( b,A\right) $ for all $B\subseteq A$ in $\mathcal{%
A}$ and all $a,b\in B$;

\item[(iv)] $p$ satisfies Odds Independence;

\item[(v)] $p$ is such that $p\left( Y\cap B,A\right) =p\left( Y,B\right)
p\left( B,A\right) $ for all $B\subseteq A$ in $\mathcal{A}$ and all $%
Y\subseteq X$.
\end{enumerate}

Moreover, in this case, $p$ satisfies Positivity if and only if it satisfies
Full Support.
\end{lemma}

\noindent \textbf{Proof} \emph{(i) implies (ii).} Choose as $C$ the
singleton $a$ appearing in the statement of the axiom.

\emph{(ii) implies (iii).} Given any $B\subseteq A$ in $\mathcal{A}$ and any 
$a,b\in B$, by the Choice Axiom, $p\left( a,A\right) =p\left( a,B\right)
p\left( B,A\right) $, but then $p\left( b,B\right) p\left( a,A\right)
=p\left( a,B\right) p\left( b,B\right) p\left( B,A\right) =p\left(
a,B\right) p\left( b,A\right) $ where the second equality follows from
another application of the Choice Axiom.

\emph{(iii) implies (iv).} Let $A\in \mathcal{A}$ and arbitrarily choose $%
a,b\in A$ such that $p\left( a,A\right) /p\left( b,A\right) \neq 0/0$. By
(iii),%
\begin{equation*}
p\left( b,a\right) p\left( a,A\right) =p\left( b,\left \{ a,b\right \}
\right) p\left( a,A\right) =p\left( a,\left \{ a,b\right \} \right) p\left(
b,A\right) =p\left( a,b\right) p\left( b,A\right)
\end{equation*}%
three cases have to be considered:

\begin{itemize}
\item $p\left( b,a\right) \neq0$ and $p\left( b,A\right) \neq0$, then $%
p\left( a,A\right) /p\left( b,A\right) =p\left( a,b\right) /p\left(
b,a\right) $;

\item $p\left( b,a\right) =0$, then $p\left( a,b\right) p\left( b,A\right)
=0 $, but $p\left( a,b\right) \neq0$ (because $p\left( a,b\right) /p\left(
b,a\right) \neq0/0$), thus $p\left( b,A\right) =0$ and $p\left( a,A\right) $ 
$\neq0$ (because $p\left( a,A\right) /p\left( b,A\right) \neq0/0$); therefore%
\begin{equation*}
\frac{p\left( a,b\right) }{p\left( b,a\right) }=\infty=\frac{p\left(
a,A\right) }{p\left( b,A\right) }
\end{equation*}

\item $p\left( b,A\right) =0$, then $p\left( b,a\right) p\left( a,A\right)
=0 $, but $p\left( a,A\right) \neq0$ (because $p\left( a,A\right) /p\left(
b,A\right) \neq0/0$), thus $p\left( b,a\right) =0$ and $p\left( a,b\right) $ 
$\neq0$ (because $p\left( a,b\right) /p\left( b,a\right) \neq0/0$); therefore%
\begin{equation*}
\frac{p\left( a,A\right) }{p\left( b,A\right) }=\infty=\frac{p\left(
a,b\right) }{p\left( b,a\right) }
\end{equation*}
\end{itemize}

\emph{(iv) implies (iii).} Given any $B\subseteq A$ in $\mathcal{A}$ and any 
$a,b\in B$:

\begin{itemize}
\item If $p\left( a,A\right) /p\left( b,A\right) \neq 0/0$ and $p\left(
a,B\right) /p\left( b,B\right) \neq 0/0$, then by (\ref{IIA})%
\begin{equation*}
\frac{p\left( a,A\right) }{p\left( b,A\right) }=\frac{p\left( a,b\right) }{%
p\left( b,a\right) }=\frac{p\left( a,B\right) }{p\left( b,B\right) }
\end{equation*}

\begin{itemize}
\item[$\circ$] If $p\left( b,A\right) \neq0$, then $p\left( b,B\right) \neq0$
and $p\left( b,B\right) p\left( a,A\right) =p\left( a,B\right) p\left(
b,A\right) $.

\item[$\circ$] Else $p\left( b,A\right) =0$, then $p\left( b,B\right) =0$
and again $p\left( b,B\right) p\left( a,A\right) =p\left( a,B\right) p\left(
b,A\right) $.
\end{itemize}

\item Else, either $p\left( a,A\right) /p\left( b,A\right) =0/0$ or $p\left(
a,B\right) /p\left( b,B\right) =0/0$, and in both cases 
\begin{equation*}
p\left( b,B\right) p\left( a,A\right) =p\left( a,B\right) p\left( b,A\right)
\end{equation*}
\end{itemize}

\emph{(iii) implies (v).} Given any $B\subseteq A$ in $\mathcal{A}$ and any $%
Y\subseteq X$, since $p\left( B,B\right) =1$, it follows $p\left( Y,B\right)
=p\left( Y\cap B,B\right) $. Therefore 
\begin{align*}
p\left( Y\cap B,A\right) & =\sum_{y\in Y\cap B}p\left( y,A\right)
=\sum_{y\in Y\cap B}\left( \sum_{x\in B}p\left( x,B\right) \right) p\left(
y,A\right) =\sum_{y\in Y\cap B}\left( \sum_{x\in B}p\left( x,B\right)
p\left( y,A\right) \right) \\
\text{\lbrack by (iii)]}& =\sum_{y\in Y\cap B}\left( \sum_{x\in B}p\left(
y,B\right) p\left( x,A\right) \right) =\sum_{y\in Y\cap B}p\left( y,B\right)
\left( \sum_{x\in B}p\left( x,A\right) \right) \\
& =\sum_{y\in Y\cap B}p\left( y,B\right) p\left( B,A\right) =p\left( Y\cap
B,B\right) p\left( B,A\right) =p\left( Y,B\right) p\left( B,A\right)
\end{align*}

\emph{(v) implies (i).} Take $Y=C$.

Finally, let $p$ satisfy the Choice Axiom. Assume -- \emph{per contra} --
Positivity holds and $p\left( a,A\right) =0$ for some $A\in \mathcal{A}$ and
some $a\in A$. Then $A\neq \left \{ a\right \} $ and, for all $b\in
A\setminus \left \{ a\right \} $, the Choice Axiom implies $0=p\left(
a,A\right) =p\left( a,\left \{ a,b\right \} \right) p\left( \left \{
a,b\right \} ,A\right) =p\left( a,b\right) \left( p\left( a,A\right)
+p\left( b,A\right) \right) =p\left( a,b\right) p\left( b,A\right) $ whence $%
p\left( b,A\right) =0$ (because $p\left( a,b\right) \neq0$), contradicting $%
p\left( A,A\right) =1$. Therefore Positivity implies Full Support. The
converse is trivial.\hfill$\blacksquare \bigskip$

If $p:\mathcal{A}\rightarrow \Delta \left( X\right) $ is a random choice
rule, denote by $\sigma_{p}\left( A\right) $ the support of $p_{A}$, for all 
$A\in \mathcal{A}$.

\begin{lemma}
\label{lm:warp-bis}If $p:\mathcal{A}\rightarrow \Delta \left( X\right) $ is
a random choice rule that satisfies the Choice Axiom, then $\sigma _{p}:%
\mathcal{A}\rightarrow \mathcal{A}$ is a rational choice correspondence.
\end{lemma}

\noindent \textbf{Proof }Clearly, $\varnothing \neq \sigma_{p}\left(
A\right) \subseteq A$ for all $A\in \mathcal{A}$, then $\sigma_{p}:\mathcal{A%
}\rightarrow \mathcal{A}$ is a choice correspondence. Let $A,B\in \mathcal{A}
$ be such that\ $B\subseteq A$ and assume that $\sigma_{p}\left( A\right)
\cap B\neq \varnothing$.

We want to show that $\sigma_{p}\left( A\right) \cap B=\sigma_{p}\left(
B\right) $. Since $p$ satisfies the Choice Axiom, if $a\in \sigma_{p}\left(
A\right) \cap B$, then $0<p\left( a,A\right) =p\left( a,B\right) p\left(
B,A\right) $. It follows that $p\left( a,B\right) >0$, that is, $a\in
\sigma_{p}\left( B\right) $. Thus, $\sigma_{p}\left( A\right) \cap
B\subseteq \sigma_{p}\left( B\right) $. As to the converse inclusion, let $%
a\in \sigma_{p}\left( B\right) $, that is, $p\left( a,B\right) >0$.\ By
contradiction, assume that $a\notin \sigma_{p}\left( A\right) \cap B$. Since 
$a\in B$, it must be the case that $a\notin \sigma_{p}\left( A\right) $,
that is, $p\left( a,A\right) =0$. Since $p$ satisfies the Choice Axiom, we
then have $0=p\left( a,A\right) =p\left( a,B\right) p\left( B,A\right) $.
Since $p\left( a,B\right) >0$,\ it must be the case that $p\left( B,A\right)
=0$, that is, $\sigma_{p}\left( A\right) \cap B=\varnothing$. This
contradicts $\sigma_{p}\left( A\right) \cap B\neq \varnothing$; therefore, $%
a $ belongs to $\sigma_{p}\left( A\right) \cap B$. Thus, $\sigma_{p}\left(
B\right) \subseteq \sigma_{p}\left( A\right) \cap B$.\hfill$\blacksquare$

\begin{lemma}
\label{lem:warp-bis}The following conditions are equivalent for a function $%
p:\mathcal{A}\rightarrow \Delta \left( X\right) $:

\begin{enumerate}
\item[(i)] $p$ is a random choice rule that satisfies the Choice Axiom;

\item[(ii)] $p$ is a random choice rule such that $\sigma _{p}$ is a
rational choice correspondence, and%
\begin{equation}
p_{B}\left( a\right) =\frac{p_{H}\left( a\right) }{p_{H}\left( B\right) }
\label{eq:uellavacca}
\end{equation}%
for all $B\subseteq H\in \mathcal{A}$ and all $a\in \sigma _{p}\left(
H\right) \cap B$;

\item[(iii)] there exist a function $v:X\rightarrow \left( 0,\infty \right) $
and a\ rational choice correspondence $\Gamma :\mathcal{A}\rightarrow 
\mathcal{A}$ such that, for all $x\in X$ and $A\in \mathcal{A}$%
\begin{equation}
p\left( x,A\right) =\left \{ 
\begin{array}{ll}
\dfrac{v\left( x\right) }{\sum_{b\in \Gamma \left( A\right) }v\left(
b\right) }\medskip & \qquad \text{if }x\in \Gamma \left( A\right) \\ 
0 & \qquad \text{else}%
\end{array}%
\right.  \label{eq:rcr-opt}
\end{equation}
\end{enumerate}

In this case, $\Gamma $ is unique and coincides with $\sigma _{p}$.
\end{lemma}

\noindent \textbf{Proof} \emph{(iii) implies (i).} Let $p$ be given by (\ref%
{eq:rcr-opt}) with $\Gamma $ a rational choice correspondence and $%
v:X\rightarrow \left( 0,\infty \right) $. It is easy to check that $p$ is a
well defined random choice rule, that the support correspondence $\limfunc{%
supp}p$ coincides with $\Gamma $, and that 
\begin{equation*}
p\left( Y,A\right) =\sum_{y\in Y\cap \Gamma \left( A\right) }\dfrac{v\left(
y\right) }{\sum_{d\in \Gamma \left( A\right) }v\left( d\right) }
\end{equation*}%
for all $Y\subseteq X$ and all $A\in \mathcal{A}$.

Let $A,B\in \mathcal{A}$ be such that $B\subseteq A$ and $a\in B$. We have
two cases:

\begin{itemize}
\item If $\Gamma \left( A\right) \cap B\neq \varnothing$, since $\Gamma$
satisfies WARP,\ $\Gamma \left( A\right) \cap B=\Gamma \left( B\right) $.

\begin{itemize}
\item[$\circ $] If $a\in \Gamma \left( B\right) $, then $a\in \Gamma \left(
A\right) $ and $p\left( a,B\right) =v\left( a\right) /\sum_{b\in \Gamma
\left( B\right) }v\left( b\right) $, it follows that%
\begin{equation*}
p\left( a,A\right) =\frac{v\left( a\right) }{\sum_{d\in \Gamma \left(
A\right) }v\left( d\right) }=\frac{v\left( a\right) }{\sum_{b\in \Gamma
\left( B\right) }v\left( b\right) }\frac{\sum_{b\in \Gamma \left( A\right)
\cap B}v\left( b\right) }{\sum_{d\in \Gamma \left( A\right) }v\left(
d\right) }=p\left( a,B\right) p\left( B,A\right)
\end{equation*}

\item[$\circ $] Else $a\notin \Gamma \left( B\right) $, and since $a\in B$,
it must be the case that $a\notin \Gamma \left( A\right) $, so $p\left(
a,A\right) =0=p\left( a,B\right) =p\left( a,B\right) p\left( B,A\right) $.
\end{itemize}

\item Else $\Gamma \left( A\right) \cap B=\varnothing$.\ It follows that\ $%
a\notin \Gamma \left( A\right) $ and\ $p\left( B,A\right) =0=p\left(
a,A\right) $; again, we have $p\left( a,A\right) =p\left( a,B\right) p\left(
B,A\right) $.
\end{itemize}

These cases prove that $p$\ satisfies the Choice Axiom.

\emph{(i) implies (ii).}\ Let $p:\mathcal{A}\rightarrow \Delta \left(
X\right) $ be a random choice rule that satisfies the Choice Axiom. Then, by
Lemma \ref{lm:warp-bis}, $\sigma _{p}:\mathcal{A}\rightarrow \mathcal{A}$ is
a rational choice correspondence. Moreover, if $B\subseteq H$ and all $a\in
\sigma _{p}\left( H\right) \cap B$, then 
\begin{equation*}
p\left( a,H\right) =p\left( a,B\right) p\left( B,H\right)
\end{equation*}%
but $p\left( B,H\right) \geq p\left( a,H\right) >0$ because $a\in B$ and $%
a\in \sigma _{p}\left( H\right) $, and (\ref{eq:uellavacca}) follows.

\emph{(ii) implies (iii).}\ Let $p:\mathcal{A}\rightarrow \Delta \left(
X\right) $ be a random choice rule such that $\sigma _{p}$ is a rational
choice correspondence, and that satisfies (\ref{eq:uellavacca}). Since, $%
\sigma _{p}$ is a rational choice correspondence, then the relation 
\begin{equation*}
a\succsim b\iff a\in \sigma _{p}\left( \left \{ a,b\right \} \right) \iff
p\left( a,b\right) >0
\end{equation*}%
is a weak order on $X$; and its symmetric part $\sim $ is an equivalence
relation such that%
\begin{equation*}
a\sim b\iff p\left( a,b\right) >0\text{ and }p\left( b,a\right) >0\iff
r\left( a,b\right) \in \left( 0,\infty \right)
\end{equation*}%
Moreover, by Theorem 3 of Arrow (1959), it follows that%
\begin{equation}
\sigma _{p}\left( A\right) =\left \{ a\in A:a\succsim b\quad \forall b\in
A\right \} \qquad \forall A\in \mathcal{A}  \label{eq:arg-max-fin}
\end{equation}%
in particular, all elements of $\sigma _{p}\left( A\right) $ are equivalent
with respect to $\sim $, and 
\begin{equation}
\sigma _{p}\left( S\right) =S  \label{eq:arg-max-ind}
\end{equation}%
for all $S\in \mathcal{A}$ consisting of equivalent elements.

Let $\left \{ X_{i}:i\in I\right \} $ be the family of all equivalence
classes of $\sim $ in $X$. Choose $a_{i}\in X_{i}$ for all $i\in I$. For
each $x\in X $, there exists one and only one $i=i_{x}$ such that $x\in
X_{i} $, set%
\begin{equation}
v\left( x\right) =r\left( x,a_{i}\right)  \label{eq:phi}
\end{equation}%
Since $x\sim a_{i}$, then $r\left( x,a_{i}\right) \in \left( 0,\infty
\right) $; and so $v:X\rightarrow \left( 0,\infty \right) $ is well defined.
Consider any $x\sim y$ in $X$ and any $S\in \mathcal{A}$ consisting of
equivalent elements and containing $x$ and $y$. Notice that, by (\ref%
{eq:arg-max-ind}), $\sigma _{p}\left( S\right) =S$, hence $x\in \sigma
_{p}\left( S\right) \cap \left \{ x,y\right \} $, then by (\ref%
{eq:uellavacca}) with $H=S$ and $B=\left \{ x,y\right \} $,%
\begin{equation*}
p\left( x,y\right) =\frac{p_{S}\left( x\right) }{p_{S}\left( \left \{
x,y\right \} \right) }
\end{equation*}%
therefore%
\begin{equation*}
0<p\left( x,S\right) =p\left( x,y\right) p\left( \left \{ x,y\right \}
,S\right)
\end{equation*}%
and analogously%
\begin{equation*}
0<p\left( y,S\right) =p\left( y,x\right) p\left( \left \{ x,y\right \}
,S\right)
\end{equation*}%
yielding that%
\begin{equation}
p\left( x,y\right) p\left( y,x\right) p\left( x,S\right) p\left( y,S\right)
>0\text{ and }\frac{p\left( x,S\right) }{p\left( y,S\right) }=\frac{p\left(
x,y\right) }{p\left( y,x\right) }=r\left( x,y\right)  \label{eq:S}
\end{equation}%
We are ready to conclude our proof, that is, to show that (\ref{eq:rcr-opt})
holds with $\Gamma =\sigma _{p}$. Let $a\in X$ and $A\in \mathcal{A}$. If $%
a\notin \sigma _{p}\left( A\right) $, then $p\left( a,A\right) =0$ because $%
\sigma _{p}\left( A\right) $ is the support of $p_{A}$. Else, $a\in \sigma
_{p}\left( A\right) $, and, by (\ref{eq:arg-max-fin}), all the elements in $%
\sigma _{p}\left( A\right) $ are equivalent with respect to $\sim $ and
therefore they are equivalent to some $a_{i}$ with $i\in I$. It follows that 
$\sigma _{p}\left( A\right) \cup \left \{ a_{i}\right \} \in \mathcal{A}$
and it is such that $\sigma _{p}\left( A\right) \cup \left \{ a_{i}\right \}
\subseteq X_{i}$. By (\ref{eq:arg-max-ind}), we have that $\sigma _{p}\left(
\sigma _{p}\left( A\right) \cup \left \{ a_{i}\right \} \right) =\sigma
_{p}\left( A\right) \cup \left \{ a_{i}\right \} $, that is, $p\left(
x,\sigma _{p}\left( A\right) \cup \left \{ a_{i}\right \} \right) >0$ for
all $x\in \sigma _{p}\left( A\right) \cup \left \{ a_{i}\right \} $ and $%
p\left( \sigma _{p}\left( A\right) ,\sigma _{p}\left( A\right) \cup \left \{
a_{i}\right \} \right) >0$. By (\ref{eq:uellavacca}) with $H=A$ and $%
B=\sigma _{p}\left( A\right) $, since $a\in \sigma _{p}\left( A\right) \cap
B $, it follows%
\begin{equation*}
p\left( a,\sigma _{p}\left( A\right) \right) =\frac{p\left( a,A\right) }{%
p\left( \sigma _{p}\left( A\right) ,A\right) }
\end{equation*}%
Since $p\left( \sigma _{p}\left( A\right) ,A\right) =1$, then%
\begin{equation*}
p\left( a,A\right) =p\left( a,\sigma _{p}\left( A\right) \right)
\end{equation*}%
By (\ref{eq:uellavacca}) again, with $H=\sigma _{p}\left( A\right) \cup
\left \{ a_{i}\right \} $ and $B=\sigma _{p}\left( A\right) $, since $a\in
\sigma _{p}\left( \sigma _{p}\left( A\right) \cup \left \{ a_{i}\right \}
\right) \cap \sigma _{p}\left( A\right) $, then%
\begin{equation*}
p\left( a,\sigma _{p}\left( A\right) \right) =\frac{p\left( a,\sigma
_{p}\left( A\right) \cup \left \{ a_{i}\right \} \right) }{p\left( \sigma
_{p}\left( A\right) ,\sigma _{p}\left( A\right) \cup \left \{ a_{i}\right \}
\right) }=\frac{\frac{p\left( a,\sigma _{p}\left( A\right) \cup \left \{
a_{i}\right \} \right) }{p\left( a_{i},\sigma _{p}\left( A\right) \cup \left
\{ a_{i}\right \} \right) }}{\frac{p\left( \sigma _{p}\left( A\right)
,\sigma _{p}\left( A\right) \cup \left \{ a_{i}\right \} \right) }{p\left(
a_{i},\sigma _{p}\left( A\right) \cup \left \{ a_{i}\right \} \right) }}
\end{equation*}%
applying (\ref{eq:S}) to the pairs $\left( x,y\right) =\left( a,a_{i}\right) 
$ and $\left( x,y\right) =\left( b,a_{i}\right) $, with $b\in \sigma
_{p}\left( A\right) $, in $S=\sigma _{p}\left( A\right) \cup \left \{
a_{i}\right \} \subseteq X_{i}$, we can conclude that%
\begin{equation*}
\frac{\frac{p\left( a,\sigma _{p}\left( A\right) \cup \left \{ a_{i}\right
\} \right) }{p\left( a_{i},\sigma _{p}\left( A\right) \cup \left \{
a_{i}\right \} \right) }}{\frac{p\left( \sigma _{p}\left( A\right) ,\sigma
_{p}\left( A\right) \cup \left \{ a_{i}\right \} \right) }{p\left(
a_{i},\sigma _{p}\left( A\right) \cup \left \{ a_{i}\right \} \right) }}=%
\frac{\frac{p\left( a,\sigma _{p}\left( A\right) \cup \left \{ a_{i}\right
\} \right) }{p\left( a_{i},\sigma _{p}\left( A\right) \cup \left \{
a_{i}\right \} \right) }}{\sum_{b\in \sigma _{p}\left( A\right) }\frac{%
p\left( b,\sigma _{p}\left( A\right) \cup \left \{ a_{i}\right \} \right) }{%
p\left( a_{i},\sigma _{p}\left( A\right) \cup \left \{ a_{i}\right \}
\right) }}=\frac{r\left( a,a_{i}\right) }{\sum_{b\in \sigma _{p}\left(
A\right) }r\left( b,a_{i}\right) }=\frac{v\left( a\right) }{\sum_{b\in
\sigma _{p}\left( A\right) }v\left( b\right) }
\end{equation*}%
as wanted.

As for the uniqueness part, we already observed that (iii) implies $\Gamma
=\sigma _{p}$.\hfill $\blacksquare \bigskip $

\textbf{Theorem \ref{thm:main}} immediately follows.\bigskip

\noindent \textbf{Proof of Proposition \ref{prop:u}} In Theorem \ref%
{thm:main}, $\Gamma $ is a rational choice correspondence and the
corresponding weak order is%
\begin{equation*}
a\succsim b\iff a\in \Gamma \left( \left \{ a,b\right \} \right) \iff
p\left( a,b\right) >0
\end{equation*}%
thus Continuity can be rewritten as:\emph{\ }Given any $x,y\in X$,\ if $%
\left \{ x_{n}\right \} _{n\in \mathbb{N}}$\ converges to $x$, then%
\begin{eqnarray*}
x_{n} &\succsim &y\text{\ for all }n\in \mathbb{N}\text{ }\implies x\succsim
y \\
y &\succsim &x_{n}\text{\ for all }n\in \mathbb{N}\text{ }\implies y\succsim
x
\end{eqnarray*}%
This concludes the proof, because on a separable metric space, a weak order
admits a continuous utility if and only if its upper and lower level sets
are closed (see, e.g., Kreps, 1988, p. 27).\hfill $\blacksquare $\bigskip

The set $\mathcal{W}$ of all weak orders on $X$ is endowed with the $\sigma $%
-algebra $\mathfrak{W}$ generated by the sets of the form%
\begin{equation*}
W_{ab}=\left \{ \succ \ :\ a\succ b\right \} \qquad \forall a,b\in X
\end{equation*}

Given $\succ $ and $\succ ^{\prime }$ in $\mathcal{W}$, the lexicographic
composition $\succ \circ \succ ^{\prime }$ of $\succ $ and $\succ ^{\prime }$
is routinely seen to be a weak order too (see, e.g., Fishburn, 1974).

\begin{lemma}
For each $\succ $ in $\mathcal{W}$, the map%
\begin{equation*}
\begin{array}{llll}
f=f_{\succ }: & \mathcal{W} & \rightarrow & \mathcal{W} \\ 
& \succ ^{\prime } & \mapsto & \succ \circ \succ ^{\prime }%
\end{array}%
\end{equation*}%
is measurable with respect to $\mathfrak{W}$.
\end{lemma}

\noindent \textbf{Proof} Arbitrarily choose $a,b\in X$, and study%
\begin{eqnarray*}
f^{-1}\left( W_{ab}\right) &=&f^{-1}\left( \left \{ \succ ^{\prime \prime }\
:\ a\succ ^{\prime \prime }b\right \} \right) =\left \{ \succ ^{\prime }\ :\
f\left( \succ ^{\prime }\right) \in \left \{ \succ ^{\prime \prime }\ :\
a\succ ^{\prime \prime }b\right \} \right \} \\
&=&\left \{ \succ ^{\prime }\ :\ a\ f\left( \succ ^{\prime }\right) \
b\right \} =\left \{ \succ ^{\prime }\ :\ a\succ \circ \succ ^{\prime
}b\right \}
\end{eqnarray*}

\begin{itemize}
\item if $a\prec b$, then there is no $\succ ^{\prime }$ in $\mathcal{W}$
such that $a\succ \circ \succ ^{\prime }b$, that is,%
\begin{equation*}
\left \{ \succ ^{\prime }\ :\ a\succ \circ \succ ^{\prime }b\right \}
=\varnothing
\end{equation*}%
which is measurable (because $\varnothing \in \mathfrak{W}$),

\item else if $a\succ b$, then $a\succ \circ \succ ^{\prime }b$ for all $%
\succ ^{\prime }$ in $\mathcal{W}$, that is,%
\begin{equation*}
\left \{ \succ ^{\prime }\ :\ a\succ \circ \succ ^{\prime }b\right \} =%
\mathcal{W}
\end{equation*}%
which is measurable (because $\mathcal{W}\in \mathfrak{W}$),

\item else, it must be the case that $a\sim b$ and $a\succ \circ \succ
^{\prime }b$ if and only if $a\succ ^{\prime }b$, that is,%
\begin{equation*}
\left \{ \succ ^{\prime }\ :\ a\succ \circ \succ ^{\prime }b\right \} =\left
\{ \succ ^{\prime }\ :\ a\succ ^{\prime }b\right \} =W_{ab}
\end{equation*}%
which is measurable (because $W_{ab}\in \mathfrak{W}$).
\end{itemize}

Therefore $f$ is measurable since the counterimage of a class of generators
of $\mathfrak{W}$ is contained in $\mathfrak{W}$.\hfill $\blacksquare
\bigskip $

A \emph{Random Preference Model} is a measurable function%
\begin{equation*}
\begin{array}{cccc}
P: & \left( \Omega ,\mathcal{F},\Pr \right) & \rightarrow & \mathcal{W} \\ 
& \omega & \mapsto & P\left( \omega \right)%
\end{array}%
\end{equation*}%
It is common practice to write $\succ _{\omega }$\ instead of $P\left(
\omega \right) $. The \emph{Random Selector} $p$ based on the RPM $P$ is
given by 
\begin{equation*}
p\left( a,A\right) =\Pr \left( \omega \in \Omega :a\succ _{\omega }b\text{%
\quad }\forall b\in A\setminus \left \{ a\right \} \right) \qquad \forall
a\in A\in \mathcal{A}
\end{equation*}%
The latter is well defined because%
\begin{eqnarray*}
\left \{ \omega \in \Omega :a\succ _{\omega }b\text{\quad }\forall b\in
A\setminus \left \{ a\right \} \right \} &=&\left \{ \omega \in \Omega
:P\left( \omega \right) \in W_{ab}\text{\quad }\forall b\in A\setminus \left
\{ a\right \} \right \} \\
&=&\left \{ \omega \in \Omega :P\left( \omega \right) \in \bigcap
\nolimits_{b\in A\setminus \left \{ a\right \} }W_{ab}\right \} \\
&=&P^{-1}\left( \bigcap \nolimits_{b\in A\setminus \left \{ a\right \}
}W_{ab}\right) \in \mathcal{F}
\end{eqnarray*}%
since $P$ is measurable. Moreover, depending on $P$, the RS $p$ might not
define a random choice rule. For instance, if $P$ is constantly equal to the
trivial weak order according to which all alternatives are indifferent, then 
$p\left( a,A\right) =0$\ for all $a\in A\in \mathcal{A}$ such that $%
\left
\vert A\right \vert \geq 2$.\bigskip

The proof of Proposition \ref{prop:lexi} hinges on the study of the
composition of the functions $f_{\succ }$ and $P$.

First, such a composition defines a random preference model, because%
\begin{equation*}
\begin{array}{llll}
f_{\succ }\circ P: & \left( \Omega ,\mathcal{F},\Pr \right) & \rightarrow & 
\mathcal{W} \\ 
& \omega & \mapsto & f_{\succ }\left( P\left( \omega \right) \right) =~\succ
\circ \succ _{\omega }%
\end{array}%
\end{equation*}%
---being a composition of measurable functions, it is measurable.

Second, the random selector based on the random preference model $f_{\succ
}\circ P$ is a lexicographic version of $P$, that first selects the
maximizers of $\succ $, then breaks the ties according to $P$.

In order to state these results formally, we denote by $\Gamma =\Gamma
_{\succ }$ the rational choice correspondence induced by $\succ $.\footnote{$%
\Gamma _{\succ }\left( A\right) =\left \{ a\in A:a\succsim b\text{ for all }%
b\in A\right \} $ also recall that $a\succsim b$ if and only if $a\nprec b$.}

\begin{lemma}
\label{lem:newlemma}Let $\succ $ be a weak order, $P=\left \{ \succ _{\omega
}\right \} _{\omega \in \Omega }$ be a RPM, and $p$ be the RS based on $P$.
Then $f_{\succ }\circ P=\left \{ \succ \circ \succ _{\omega }\right \}
_{\omega \in \Omega }$ is a RPM and the RS based on it is given by 
\begin{equation}
p_{\succ }\left( a,A\right) =\left \{ 
\begin{array}{ll}
p\left( a,\Gamma \left( A\right) \right) \medskip & \qquad \text{if }a\in
\Gamma \left( A\right) \\ 
0 & \qquad \text{else}%
\end{array}%
\right.  \label{ex:lexi-rs}
\end{equation}%
for all $a\in A\in \mathcal{A}$.
\end{lemma}

\noindent \textbf{Proof} We already observed that $f_{\succ }\circ
P=\left
\{ \succ \circ \succ _{\omega }\right \} _{\omega \in \Omega }$ is
a RPM. By definition of random selector based on a RPM%
\begin{equation*}
p_{\succ }\left( a,A\right) =\Pr \left( \omega \in \Omega :a\succ \circ
\succ _{\omega }b\text{\quad }\forall b\in A\setminus \left \{ a\right \}
\right)
\end{equation*}%
We have to verify that this formula coincides with (\ref{ex:lexi-rs}) for
all $a\in A\in \mathcal{A}$.

For each $A\in \mathcal{A}$ and each $a\in \Gamma \left( A\right) $, set 
\begin{eqnarray*}
J_{A}\left( a\right) &=&\left \{ \omega \in \Omega :a\succ _{\omega }c\text{
for all }c\in \Gamma \left( A\right) \setminus \left \{ a\right \} \right \}
=J \\
K_{A}\left( a\right) &=&\left \{ \omega \in \Omega :a\succ \circ \succ
_{\omega }b\text{ for all }b\in A\setminus \left \{ a\right \} \right \} =K
\end{eqnarray*}%
Next we check that $J=K$.

If $\omega \in J$, then $a\succ _{\omega }c$ for all $c\in \Gamma \left(
A\right) \setminus \left \{ a\right \} $; take any $b\in A\setminus \left \{
a\right \} $,

\begin{itemize}
\item if $b$ is such that $b\notin \Gamma \left( A\right) $, then, $a\succ b$
and hence $a\succ \circ \succ _{\omega }b$,

\item else $b\in \Gamma \left( A\right) $, then $a\sim b$ and $a\succ
_{\omega }b$, again $a\succ \circ \succ _{\omega }b$,
\end{itemize}

\noindent then $a\succ \circ \succ _{\omega }b$ for all $b\in A\setminus
\left \{ a\right \} $, thus $\omega \in K$.

Conversely, if $\omega \in K$, then $a\succ \circ \succ _{\omega }b$ for all 
$b\in A\setminus \left \{ a\right \} $. Thus, for all $b\in \Gamma \left(
A\right) \setminus \left \{ a\right \} $, since relation $a\sim b$, it must
be the case that $a\succ _{\omega }b$. Therefore $\omega $ is such that $%
a\succ _{\omega }b$ for all $b\in \Gamma \left( A\right) \setminus \left \{
a\right \} $, and $\omega \in J$.

Summing up, for all $A\in \mathcal{A}$ and $a\in \Gamma \left( A\right) $, 
\begin{equation*}
p\left( a,\Gamma \left( A\right) \right) =\Pr J_{A}\left( a\right) =\Pr
K_{A}\left( a\right) =p_{\succ }\left( a,A\right)
\end{equation*}%
and the first line of (\ref{ex:lexi-rs}) is true.

Let $A\in \mathcal{A}$ and $a\notin \Gamma \left( A\right) $, then there
exists $\bar{b}\in A\setminus \left \{ a\right \} $ such that $a\prec \bar{b}
$, and for no $\omega $ it holds $a\succ \circ \succ _{\omega }\bar{b}$,
that is,%
\begin{equation*}
K_{A}\left( a\right) =\left \{ \omega \in \Omega :a\succ \circ \succ
_{\omega }b\text{ for all }b\in A\setminus \left \{ a\right \} \right \}
=\varnothing
\end{equation*}%
therefore $p_{\succ }\left( a,A\right) =\Pr K_{A}\left( a\right) =0$, and
the second line of (\ref{ex:lexi-rs}) is true too.\hfill $\blacksquare
\bigskip $

\noindent \textbf{Proof of Proposition \ref{prop:lexi}} The equivalence
between points (i) and (ii) corresponds with the equivalence between the
points with the same name of Lemma \ref{lem:warp-bis}.

(i) implies (iii). By Theorem \ref{thm:main}, there exist a function $\alpha
:X\rightarrow \mathbb{R}$ and a rational choice correspondence $\Gamma :%
\mathcal{A}\rightarrow \mathcal{A}$ such that 
\begin{equation*}
p\left( a,A\right) =\left \{ 
\begin{array}{ll}
\dfrac{e^{\alpha \left( a\right) }}{\sum_{b\in \Gamma \left( A\right)
}e^{\alpha \left( b\right) }}\medskip & \qquad \text{if }a\in \Gamma \left(
A\right) \\ 
0 & \qquad \text{else}%
\end{array}%
\right.
\end{equation*}%
for all $a\in A\in \mathcal{A}$. Denote by $\succ $ the weak order that
corresponds to $\Gamma $.

As shown by McFadden (1973), the Lucean random choice rule 
\begin{equation*}
q\left( a,A\right) =\dfrac{e^{\alpha \left( a\right) }}{\sum_{b\in
A}e^{\alpha \left( b\right) }}\qquad \forall a\in A\in \mathcal{A}
\end{equation*}%
is based on a (Lucean) RPM $P=\left \{ \succ _{\omega }\right \} _{\omega
\in \Omega }$. By Lemma \ref{lem:newlemma}, it follows that $f_{\succ }\circ
P=\left \{ \succ \circ \succ _{\omega }\right \} _{\omega \in \Omega }$ is a
RPM and the RS based on it is given by 
\begin{equation*}
q_{\succ }\left( a,A\right) =\left \{ 
\begin{array}{ll}
q\left( a,\Gamma \left( A\right) \right) \medskip & \qquad \text{if }a\in
\Gamma \left( A\right) \\ 
0 & \qquad \text{else}%
\end{array}%
\right. =p\left( a,A\right) \qquad \forall a\in A\in \mathcal{A}
\end{equation*}%
Therefore, there exist a Lucean Random Preference Model $\left \{ \succ
_{\omega }\right \} _{\omega \in \left( \Omega ,\mathcal{F},\Pr \right) }$
and a weak order $\succ $ on $X$ such that $p$ is based on $\left \{ \succ
\circ \succ _{\omega }\right \} _{\omega \in \left( \Omega ,\mathcal{F},\Pr
\right) }$.

(iii) implies (i). If there exist a Lucean Random Preference Model $\left \{
\succ _{\omega }\right \} _{\omega \in \left( \Omega ,\mathcal{F},\Pr
\right) }$ and a weak order $\succ $ on $X$ such that $p$ is based on $%
\left
\{ \succ \circ \succ _{\omega }\right \} _{\omega \in \left( \Omega ,%
\mathcal{F},\Pr \right) }$; in particular, there exists $\alpha
:X\rightarrow \mathbb{R}$ such that 
\begin{equation*}
\Pr \left( \omega \in \Omega :a\succ _{\omega }b\text{\quad }\forall b\in
A\setminus \left \{ a\right \} \right) =\dfrac{e^{\alpha \left( a\right) }}{%
\sum_{b\in A}e^{\alpha \left( b\right) }}\qquad \forall a\in A\in \mathcal{A}
\end{equation*}%
Denoting 
\begin{equation*}
q\left( a,A\right) =\dfrac{e^{\alpha \left( a\right) }}{\sum_{b\in
A}e^{\alpha \left( b\right) }}\qquad \forall a\in A\in \mathcal{A}
\end{equation*}%
the RS based on $\left \{ \succ _{\omega }\right \} _{\omega \in \left(
\Omega ,\mathcal{F},\Pr \right) }$, by Lemma \ref{lem:newlemma}, the RS is
based on $\left \{ \succ \circ \succ _{\omega }\right \} _{\omega \in \left(
\Omega ,\mathcal{F},\Pr \right) }$ is%
\begin{eqnarray*}
q_{\succ }\left( a,A\right) &=&\left \{ 
\begin{array}{ll}
q\left( a,\Gamma \left( A\right) \right) \medskip & \qquad \text{if }a\in
\Gamma \left( A\right) \\ 
0 & \qquad \text{else}%
\end{array}%
\right. \\
&=&\left \{ 
\begin{array}{ll}
\dfrac{e^{\alpha \left( a\right) }}{\sum_{b\in \Gamma \left( A\right)
}e^{\alpha \left( b\right) }}\medskip & \qquad \text{if }a\in \Gamma \left(
A\right) \\ 
0 & \qquad \text{else}%
\end{array}%
\right.
\end{eqnarray*}%
But, by assumption (iii), $q_{\succ }$ coincides with $p$ ($p$ is based on $%
\left \{ \succ \circ \succ _{\omega }\right \} _{\omega \in \left( \Omega ,%
\mathcal{F},\Pr \right) }$), and $\Gamma $ is a rational choice
correspondence because $\succ $ is a weak order. Then Theorem \ref{thm:main}
guarantees that $p$ satisfies the Choice Axiom.\hfill $\blacksquare $

{\color{cadmiumred}}

\end{document}